\documentclass[journal]{IEEEtran}

\usepackage[cmex10]{amsmath}  
\usepackage{amssymb}
\usepackage{amsfonts}
\usepackage{bm}

\usepackage{multirow}

\usepackage{dsfont}
\usepackage[latin1]{inputenc}
\usepackage{times}

\usepackage[noadjust]{cite}
\usepackage{graphicx}

\usepackage[caption=false,font=footnotesize]{subfig}
\usepackage{stfloats}

\newtheorem{thm}{Theorem}
\newtheorem{rem}{Remark}
\newtheorem{cor}{Corollary}

\newtheorem{dfn}{Definition}

\def\mb{\mathbf}
\def\opn{\operatorname*}
\def\bs{\boldsymbol}

\def\ss#1{{\sf #1}}

\def\resc#1{\uppercase{#1}}

\def\rvec#1{\bs{\uppercase{#1}}} 
\def\rvecr#1{\bs{\lowercase{#1}}} 

\def\mat#1{\mb{\uppercase{#1}}}

\def\C{\ds{C}}

\def\pdffun{P}

\def\EspOp{\ss{E}}

\newcommand{\Esp}[2][5]{%
  \ifcase#1
     \EspOp\{ #2 \}
     \or \EspOp \bigl\{ #2 \bigr\}
     \or \EspOp \Bigl\{ #2 \Bigr\}
     \or \EspOp \biggl\{ #2 \biggr\}
     \or \EspOp \Biggl\{ #2 \Biggr\}
  \else
     \EspOp \left\{ #2  \right\}
\fi}

\newcommand{\Earg}[3][5]{%
  \ifcase#1
     \EspOp_{#3} \{ #2 \}
     \or \EspOp_{#3} \bigl\{ #2 \bigr\}
     \or \EspOp_{#3} \Bigl\{ #2 \Bigr\}
     \or \EspOp_{#3} \biggl\{ #2 \biggr\}
     \or \EspOp_{#3} \Biggl\{ #2 \Biggr\}
  \else
     \EspOp_{#3} \left\{ #2  \right\}
\fi}

\newcommand{\CEsp}[3][5]{%
  \ifcase#1
     \EspOp\{ #2 \mid #3 \}
     \or \EspOp \bigl\{ #2 \bigm\vert #3 \bigr\}
     \or \EspOp \Bigl\{ #2 \Bigm\vert #3 \Bigr\}
     \or \EspOp \biggl\{ #2 \biggm\vert #3 \biggr\}
     \or \EspOp \Biggl\{ #2 \Biggm\vert #3 \Biggr\}
  \else
     \EspOp \left\{ #2  \,\middle\vert\, #3 \right\}
\fi}

\newcommand{\Diag}[2][5]{%
  \ifcase#1
     \mb{Diag}( #2 )
     \or \mb{Diag} \bigl( #2 \bigr)
     \or \mb{Diag} \Bigl( #2 \Bigr)
     \or \mb{Diag} \biggl( #2 \biggr)
     \or \mb{Diag} \Biggl( #2 \Biggr)
  \else
     \mb{Diag} \left( #2  \right)
\fi}

\newcommand{\diag}[2][5]{%
  \ifcase#1
     \mb{diag}( #2 )
     \or \mb{diag} \bigl( #2 \bigr)
     \or \mb{diag} \Bigl( #2 \Bigr)
     \or \mb{diag} \biggl( #2 \biggr)
     \or \mb{diag} \Biggl( #2 \Biggr)
  \else
     \mb{diag} \left( #2  \right)
\fi}

\def\Tr{\ss{Tr}}

\def\log{\ss{log}}

\def\dd{\opn{d}\!}

\def\T{\ss{T}}

\def\iid{i.i.d. }
\def\etal{et al. }

\def\EM{\mat{E}}
\def\MSE#1{\EM_{\rvec{#1}}}

\def\dim{n}
\def\I{I} 

\def\mmse{\ss{mmse}}
\def\scalart{\gamma}

\def\q{q}

\def\EM{\mat{\mathcal{E}}} 
\def\MSE#1{\EM_{\rvec{#1}}}

\def\Igen#1#2{ \I \left(  #1;#2 \right)}
\def\Icond#1#2#3{ \I \left( #1;#2 | #3 \right)}

\def\rate{\ss{R}}

\def\MSEcode{ \ss{MMSE^{\C}}}
\def\MSEcoden{ \ss{MMSE^{\C_{\dim}}}}
\def\InomGen#1#2{ \frac{1}{{\dim}} \Igen{#1}{#2} }

\def\Hb#1#2{ \frac{#2}{{\dim}} \ss{h}_b \left( #1  \right)}
\def\HdC#1#2{ \frac{1}{{\dim}}\ss{h} \left( #1  | #2 \right)}
\def\d{ \ss{d}}
\def\C{ \ss{C}}

\def\snr{\ss{snr}}

\newcounter{MYtempeqncnt}

\begin{document}

\title{MMSE of ``Bad'' Codes}
\author{Ronit~Bustin~\IEEEmembership{Student Member,~IEEE,} and Shlomo~Shamai~(Shitz)~\IEEEmembership{Fellow,~IEEE}
\thanks{This work has been supported by the Israel Science Foundation (ISF) and the CORNET Consortium.}
\thanks{Ronit Bustin is supported by the Adams Fellowship Program of the Israel Academy of Sciences and Humanities}
\thanks{The authors are with the Department of Electrical Engineering, Technion-Israel Institute of Technology, Technion City, Haifa 32000, Israel (email: bustin@tx.technion.ac.il, sshlomo@ee.technion.ac.il)}
}


\maketitle








\begin{abstract}
We examine codes, over the additive Gaussian noise channel, designed for reliable communication at some specific signal-to-noise ratio (SNR) and constrained by the permitted minimum mean-square error (MMSE) at lower SNRs. The maximum possible rate is below point-to-point capacity, and hence these are non-optimal codes (alternatively referred to as ``bad'' codes).
We show that the maximum possible rate is the one attained by superposition codebooks. Moreover, the MMSE and mutual information behavior as a function of SNR, for any code attaining the maximum rate under the MMSE constraint, is known for all SNR. We also provide a lower bound on the MMSE for finite length codes, as a function of the error probability of the code.
\end{abstract}


\section{Introduction} \label{sec:introduction}


\IEEEPARstart{C}{apacity}
and capacity achieving codes have been the main concern of information theory from the very beginning. Trying to design capacity achieving codes is a central goal of many researchers in this field. Specifically, in point-to-point channels, for which a single-letter expression of the capacity is well known \cite{CoverThomas}, the emphasis is given to the properties and design of capacity achieving codes. One such important property, derived in \cite{EXIT}, has shown that the behavior of the mutual information between the transmitted codeword and the channel output, and thus also the behavior of the minimum-mean-square error (MMSE) when estimating the transmitted codeword from the channel output, both as a function of the output's signal-to-noise ratio (SNR), of ``good'' (capacity achieving) point-to-point codes are known exactly, with no regards to the specific structure of the code.

Recently some emphasis has been given to the research of non-capacity achieving point-to-point codes \cite{Bennatan1,Bennatan2}.
These codes, referred to as ``bad'' point-to-point codes \cite{Bennatan2}, are heavily used in many multi-terminal wireless networks, and perform better, in terms of achievable rates, compared to point-to-point capacity achieving codes. Bennatan \emph{et. al.} \cite{Bennatan1} have argued that such codes have inherent benefits that often make them better candidates for multi-terminal wireless communication. For example, in \cite{EXIT} it was concluded, through the investigation of the extrinsic information (EXIT) behavior, that ``good'' codes can not function well as turbo component codes, within an iterative belief-propagation decoding procedure.

The first question that comes to mind is: What are these inherent benefits that make these codes better candidates for multi-terminal wireless communication? It is known that ``bad'' codes can obtain lower MMSE at low SNRs as compared to ``good'' point-to-point codes \cite{Bennatan2}. The hypothesis is that this property is the inherent benefit of ``bad'' codes. Surely, lower MMSE at lower SNRs is meaningless in point-to-point communication, where all that matters is the performance at the intended receiver.
However, in multi-terminal wireless networks, such as cellular networks, the case is different. In such networks there are two fundamental phenomena: interference from one node to another (an interference channel), and the potential cooperation between nodes (a relay channel). In the interference channel, where a message sent to an intended receiver acts as interference to other receivers in the network, a lower MMSE implies better possible interference cancelation, and thus improved rates for the interfered user. In the relay channel, the goal of the relay is to decode the intended message, so as to assist the transmission. In this case, a lower MMSE assist when full decoding is not possible. The relay may then use soft decoding, as suggested in \cite{Bennatan1}. These two advantages have been the center of the investigation in \cite{Bennatan1}, where two specific soft decoding algorithms, one for an interference scenario and the other for a relay scenario have been analyzed. It was shown that for ``bad'' LDPC codes, better achievable rates can be obtained, as compared to ``good'' point-to-point codes.

The problem that motivated this work is the Gaussian interference channel, where the question of how to handle interference is still open. Surely, when the interference can be decoded, as in the case of strong interference, then joint decoding is the optimal scheme and attains capacity \cite{carleialStrongInterference,SatoStrongInterference,HanKobayashi,CostaElGamal}. However, what should one do with an interference that can not be decoded. Should we treat it as noise? Should we partially decode it?
This question has been the investigation of several works. As explained above, Bennatan \etal \cite{Bennatan1} claim that soft decoding is a useful compromise in cases where complete decoding would be desirable if possible, but is not required by the terms of the problem, and show that specific ``bad'' LDPC codes attain better rates compared to ``good'' point-to-point codes. In \cite{Baccelli2011} the authors establish the capacity region of the $K$-user Gaussian interference channel, when all users are constrained to use point-to-point codes. The capacity region is shown to be achieved by a combination of treating interference as noise and joint decoding. A similar setting was also discussed in \cite{TandraInterferenceAsNoise}, and in \cite{ChaabanInterferenceAsNoise} the question whether treating interference as noise is optimal was asked on a more elaborated system of a point-to-point channel interfering with a MAC. In \cite{MotahariKhandaniInterferenceAsNoise} the authors examine the interference channel from the point of view of a single transmitter-receiver pair, being interfered. They proposed a strategy to determine the rate, by disjoining the set of interfering users into two disjoint subsets, namely the set of decodable interferences and the set of non-decodable interferences. The authors show that, when assuming that all interferences are Gaussian, their strategy achieves capacity. Finally, in \cite{AlternativeMoshksarGhasemiKhandani}, the authors examined the alternatives to treating the interference as Gaussian noise, assuming the receiver knows the constellation set used by the interferer. This makes the interference plus noise a mixed Gaussian process. Under these assumptions the authors develop an achievable rate, with improved sum-rate as compared to the one obtained when using Gaussian codebooks and treating interference as Gaussian noise.

In this work we examine a simplified scenario, as compared to the interference channel, in which we have only a single transmitter with a single intended receiver. The transmitted message reaches one or more unintended receiver, which are not interested in the transmitted message. The question asked is: if these unintended receivers wish to estimate the transmitted message with limited error, that is, some constraint on the MMSE, what is the maximum rate of transmission? The connection to the interference model is clear. Assuming that a good approach is to remove the estimated codeword, one can think of the MMSE as the remaining interference. Note that the model examined here is a simplified version as compared to the interference channel, as we have omitted the messages intended to each of the unintended receivers. However, we trust that this simplified model is an important building block towards the understanding of the interference channel, and specifically the analysis of coding schemes using partial interference cancelation.

The importance of the problem examined here is also apparent from the results obtained. We show that the optimal MMSE-wise codebook (that is, the codebook attaining the maximum rate given the MMSE constraint) in the examined setting, is the Gaussian superposition codebook. It is well known that the best achievable region for the two-user interference channel is given by the Han and Kobayashi (HK) scheme \cite{HanKobayashi}. This scheme uses partial decoding of the interfering message at the receiver. Rate splitting (that is, superposition coding) is a special case of the HK scheme, and is also point-to-point ``bad'' (see \cite[Appendix VIII-C]{Bennatan1}). It was shown in \cite{ETW} that these codes are close to optimal, and in fact are within one bit from capacity. Our results give an engineering insight to these good performance of the HK scheme.

In parallel to our work, Bandemer and El Gamal \cite{Bandemer} examined the same model but for the general discrete memoryless channel (DMC). Bandemer and El Gamal \cite{Bandemer} chose to quantify the interference (the ``disturbance'') using the mutual information at each of the unintended receivers, rather then the MMSE. They provide the rate-disturbance region: given a constraint on the disturbance, the amount of information transmitted to the unintended receiver, what is the maximum rate that can be transmitted reliably? We elaborate and compare the two methods, specifically for the Gaussian channel, in section \ref{sec:mutualInformationDisturbance}.

More specifically, we are examining the transmission of length $\dim$ codewords over a discrete memoryless standard Gaussian channel.
\begin{eqnarray} \label{eq:modelAWGN}
\resc{y} = \sqrt{ \gamma } \resc{x } + \resc{n}
\end{eqnarray}
where $\resc{n}$ is standard additive Gaussian noise. The codewords are constrained by the standard average power constraint: 
\begin{eqnarray} \label{eq:powerConstraint}
\forall \rvecr{x} \in \C_{\dim} \quad \frac{1}{n} \sum_{i = 1}^{\dim} \rvecr{x}_i^2 \leq 1
\end{eqnarray}
where $\C_{\dim}$ stands for a code of length $\dim$ codewords.

We distinguish between channel outputs at different SNRs using the following notation:
\begin{align}
Y(\gamma) = \sqrt{ \gamma } X + N
\end{align}
and for a length $\dim$ codeword we use the boldface notation:
\begin{align}
\rvec{y}(\gamma) = \sqrt{ \gamma } \rvec{x} + \rvec{n}.
\end{align}
Thus, the normalized mutual information between the input and the output will be noted as:
\begin{align}
\I_{\dim}( \gamma ) & = \InomGen{\rvec{x}}{\rvec{y}(\gamma)}.
\end{align}

The remainder of this paper is organized as follows: in section \ref{sec:Preliminary} we give some preliminary definitions and results. The problem is formulated precisely in section \ref{sec:problem}. The results are then given in the three separate sections: for a single MMSE constraint in section \ref{sec:singleMMSEConstraint}, for $K$ MMSE constraints in section \ref{sec:multiMMSEConstraints} and a lower bound on the MMSE for finite length codes is given in section \ref{sec:finiteLength}. As stated above, a comparison with the work of Bandemer and El Gamal \cite{Bandemer} is given in section \ref{sec:mutualInformationDisturbance}, adhering to an I-MMSE prespective. We conclude our work and discuss future challenges in section \ref{sec:conclusions}.

\section{Preliminary Definitions and Results} \label{sec:Preliminary}
Before formulating the problem precisely, in Section \ref{sec:problem}, we wish to define and present several key ingredients.

\subsection{Non-Optimal Code Sequences} \label{ssec:nonoptimalCodeSequences}
We begin by presenting a family of non-optimal code sequences for which our solution is valid.

\begin{dfn} \label{dfn:nonoptimalCode}
A non-optimal code-sequence $\C = \left\{ \C_{\dim} \right\}_{\dim=1}^{\infty}$, for a channel with capacity $C$, is a code-sequence with
vanishing error probability
\begin{align}
\pdffun_e^{\dim} \stackrel{\dim \to \infty}{\longrightarrow} 0 \nonumber
\end{align}
where $\pdffun_e^{\dim}$ is the error probability of the code $\C_{\dim}$, and rate satisfying
\begin{eqnarray} \label{eq:DefinitionNonOptimal}
\lim_{\dim \to \infty} \frac{1}{\dim} \log{M_{\dim}} < C.
\end{eqnarray}
where $M_{\dim}$ is the size of code $\C_{\dim}$. Moreover, we require,
\begin{align} \label{dfn:conditionMMSE}
\MSEcoden(\gamma) \stackrel{\dim \to \infty}{\longrightarrow} \MSEcode(\gamma)
\end{align}
where $\MSEcoden(\gamma) = \frac{1}{\dim} \Tr( \MSE{x}(\gamma) )$ and $\MSE{x}(\gamma)$ is the MMSE matrix defined as follows:
\begin{multline}
\MSE{x}(\gamma) = \EspOp \bigl\{ (\rvec{x} - \CEsp{\rvec{x}}{\sqrt{\gamma} \rvec{x} + \rvec{n}}) \bigr.   \\
 \bigl. (\rvec{x} - \CEsp{\rvec{x}}{\sqrt{\gamma} \rvec{x} + \rvec{n}})^\T \bigr\}
\end{multline}
with the random variable $\rvec{x}$ uniformly distributed over the $M_{\dim}$ codewords of $\C_{\dim}$.
\end{dfn}
Note that the requirement in (\ref{dfn:conditionMMSE}) is not very restrictive, as $\MSEcoden(\gamma)$ can be both upper and lower bounded by a function of $\pdffun_e^{\dim}(\gamma)$. The convergence of $\pdffun_e^{\dim}(\gamma)$ has been discussed in \cite{Feinstein}.

\subsection{The I-MMSE approach} \label{ssec:IMMSEapproach}
The approach used in order to provide insight into the MMSE constrained problem is the I-MMSE approach, this to say that we make use of the fundamental relationship between the mutual information and the MMSE in the Gaussian channel and its generalizations
\cite{IMMSE,Palomar}. Even though we are examining a scalar Gaussian channel, the $\dim$-dimensional version of this relationship is required since we are looking at the transmission of length $\dim$ codewords through the channel. In our setting the relationship is as follows:
\begin{eqnarray} \label{eq:IMMSErelationship2}
\I_n(\snr) = \frac{1}{2} \int_0^\snr \MSEcoden(\gamma) \dd \gamma .
\end{eqnarray}
Restricting our observations to the family of non-optimal code sequences defined in Definition \ref{dfn:nonoptimalCode} we can take the limit as $\dim \to \infty$ on both sides
\begin{eqnarray} \label{eq:IMMSErelationship3_infty}
\I(\snr) = \lim_{\dim \to \infty} \I_{\dim}( \snr ) = \frac{1}{2} \int_0^\snr \MSEcode(\gamma) \dd \gamma .
\end{eqnarray}
where the exchange of limit and integration on the right-hand-side is according to Lebesgue's dominated convergence theorem \cite{Williams}, the fact that $\MSEcode(\gamma)$ is upper bounded, and the condition in equation (\ref{dfn:conditionMMSE}).

The main property of the I-MMSE used in the sequel is an $\dim$-dimensional ``single crossing point'' property derived in \cite{Full} given here for completeness. This property is an extension of the scalar ``single crossing point'' property shown in \cite{PROP_full}. The following function is a simplified version (sufficient for our use in this paper) of the function defined in \cite{Full}. For an arbitrary random vector $\rvec{x}$:
\begin{IEEEeqnarray}{rCl}
\q(\rvec{x}, \sigma^2, \scalart) = \frac{\sigma^2}{1
+ \sigma^2 \scalart}  - \Tr \left(\MSE{x}(\scalart) \right). \label{eq:defqA}
\end{IEEEeqnarray}
The following theorem is proved in \cite{Full},
\begin{thm}[\cite{Full}] \label{thm:ScalarUniqueCrossingPoint}
The function $\scalart \mapsto \q(\rvec{x}, \sigma^2, \scalart)$, defined
in (\ref{eq:defqA}), has no nonnegative-to-negative zero crossings and, at most,
a single negative-to-nonnegative zero crossing in the range $\scalart \in [0,
\infty)$. Moreover, let $\snr_0 \in [0,  \infty)$ be that
negative-to-nonnegative crossing point. Then,
\begin{enumerate}
\item $\q(\rvec{x}, \sigma^2, 0) \leq 0$.
\item $\q(\rvec{x}, \sigma^2, \scalart)$ is a strictly increasing
function in the range $\scalart \in [0, \snr_0)$.
\item $\q(\rvec{x}, \sigma^2, \scalart) \geq 0$ for all $\scalart \in
[\snr_0, \infty)$.
\item $\lim_{\scalart \to \infty} \q(\rvec{x}, \sigma^2, \scalart) =
0$.
\end{enumerate}
\end{thm}
The above property is valid for all natural $\dim$, thus we may also take $\dim \to \infty$.

\subsection{Superposition Coding} \label{ssec:superposition}
\begin{figure}
\begin{center}
\setlength{\unitlength}{.1cm}
    \includegraphics[width=0.43\textwidth]{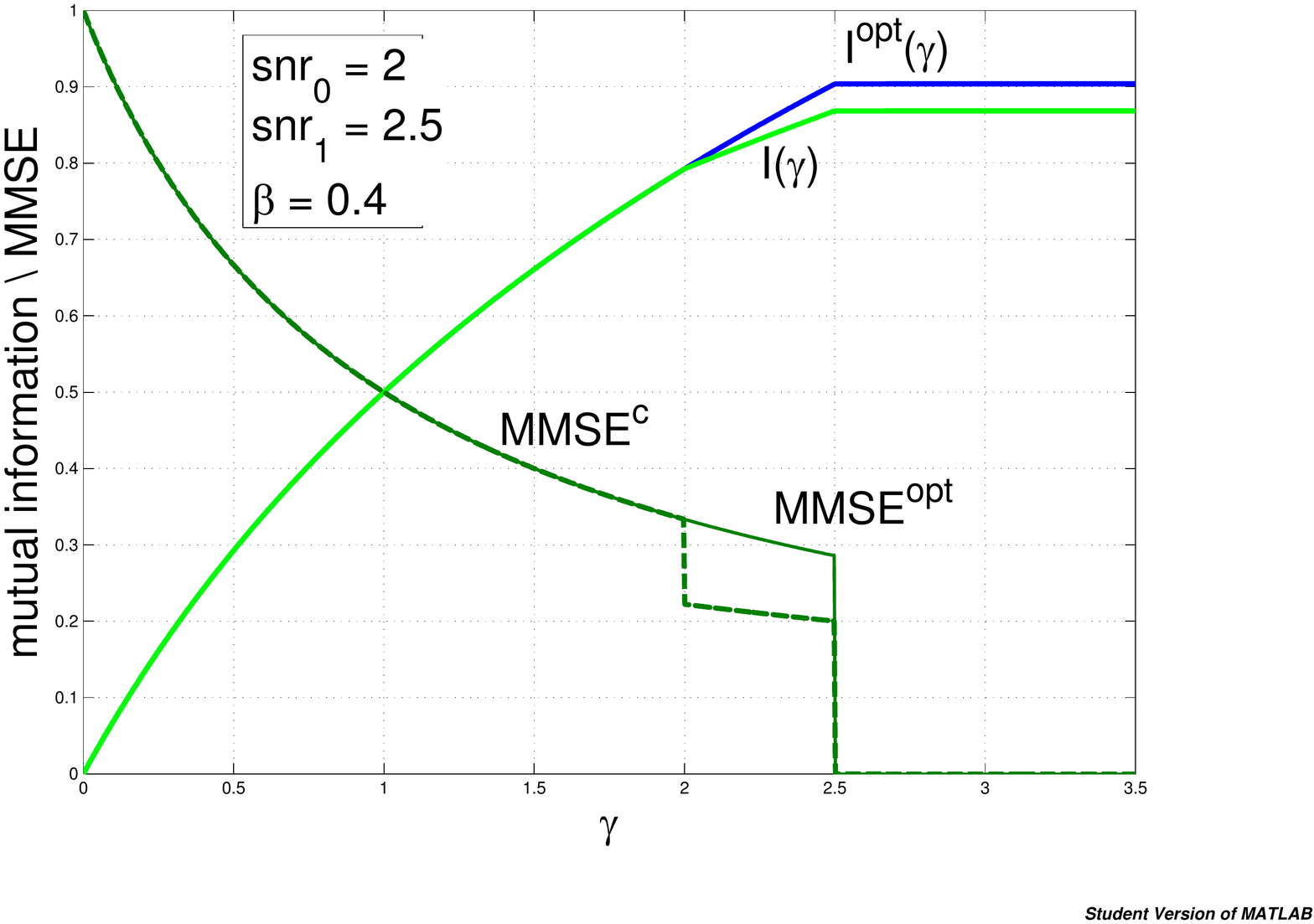}
    \caption{The mutual information and $\MSEcode(\gamma)$ of a two-layered superposition code with $(\snr_0, \snr_1) = (2,2.5)$ and $\beta = 0.4$ and the mutual information and $\MSEcode(\gamma)$ of an optimal code for rate $\snr_1$.
    } \label{fig:1}
\end{center}
\end{figure}

An important family of non-optimal codes, that is, a family of codes that do not attain the point-to-point capacity, is that of Gaussian superposition codes which are optimal for a \emph{degraded} Gaussian BC \cite{CoverThomas}. We refer to this family of codes as optimal Gaussian superposition codes. As will be shown in the sequel optimal Gaussian superposition codes are optimal MMSE-wise.
We begin by formally defining two-layered optimal Gaussian superposition code. The extension of the definition to a general $L$-layered optimal Gaussian superposition codes ($L > 1$) is straightforward.
\begin{dfn}[ \cite{CoverThomas}] \label{dfn:superposition}
Given a pair of SNRs, ($\snr_0$, $\snr_1$), where $\snr_0 < \snr_1$, two-layered optimal Gaussian superposition codes, are all codebooks that can be constructed as follows:
\begin{itemize}

\item Choose a $\beta \in (0,1)$.

\item Set $\rate_u = \frac{1}{2} \log \left( \frac{ 1 + \snr_1}{1 + \beta \snr_1} \right)$. Fill the first codebook $\C_{\dim}^u = \left\{ u_1, \cdots, u_{M_u} \right\}$ with $M_u$ \iid Gaussian vectors of average power $1 - \beta$ where $M_u = 2^{\dim \rate_u}$. This is the common message.

\item Set $\rate_v = \frac{1}{2} \log \left(  1 + \beta \snr_2 \right)$. Fill the second codebook $\C_{\dim}^v = \left\{ v_1, \cdots, v_{M_v} \right\}$ with $M_v$ \iid Gaussian vectors of average power $\beta$ where $M_v = 2^{\dim \rate_v}$. This is the private message.

\item Construct the third codebook by taking the sum $\C_{\dim} = \C_{\dim}^u + \C_{\dim}^v$, for which the cardinality is, almost surly, equal to $| \C_{\dim}^u | |\C_{\dim}^v |$. Thus, the rate is, almost surely, equal to $\frac{1}{2} \log \left( \frac{ 1 + \snr_1}{1 + \beta \snr_1} \right) + \frac{1}{2} \log \left(  1 + \beta \snr_2 \right)$.

\end{itemize}
\end{dfn}
The analysis of this family (two-layers) was done by Merhav \emph{et. al.} in \cite[section V.C]{StatisticalPhysics} from a statistical physics perspective. As noted in \cite{StatisticalPhysics}, the MMSE of this family of codebooks undergoes phase transitions, that is, it is a discontinuous function of $\gamma$.
The mutual information, $\I( \gamma)$, and $\MSEcode(\gamma)$ of this family of codebooks are known exactly and given in the next theorem (for $L = K+1$ layers).
\begin{thm}[extension of \cite{StatisticalPhysics} section V.C] \label{thm:superpositionMI_MMSE}
A $K+1$-layered optimal Gaussian superposition codebook designed for $( \snr_0, \snr_1, \cdots, \snr_K)$ with rate-splitting coefficients $\beta_0 > \cdots > \beta_{K-1}$ has the following $\I( \gamma)$:
\begin{align} \label{eq:thm_MIsuperposition}
& \left\{ \begin{array}{l}
\frac{1}{2} \log \left( 1 + \gamma \right), \quad \rm{if } \quad 0 \leq \gamma < \snr_0 \\
\frac{1}{2} \log \left( \frac{1 + \snr_0}{1 + \beta_0 \snr_0} \prod_{j=1}^{i} \frac{ 1 + \beta_{j-1} \snr_j}{1 + \beta_j \snr_j} \right) + \frac{1}{2} \log \left( 1 + \beta_i \gamma \right),  \\
\quad \rm{if } \quad \snr_i \leq \gamma \leq \snr_{i+1} \\
\frac{1}{2} \log \left( \frac{1 + \snr_0}{1 + \beta_0 \snr_0} \prod_{j=1}^{K-1} \frac{ 1 + \beta_{j-1} \snr_j}{1 + \beta_j \snr_j} \right) + \frac{1}{2} \log \left( 1 + \beta_{K-1} \snr_K \right), \\
\quad \rm{if } \quad \snr_K < \gamma
\end{array} \right.
\end{align}
and the following $\MSEcode(\gamma)$:
\begin{eqnarray} \label{eq:thm:superpositionMSE}
\MSEcode(\gamma) = \left\{ \begin{array}{ll}
\frac{1}{1 + \gamma}, & 0 \leq \gamma < \snr_0 \\
\frac{\beta_i}{1 + \beta_i \gamma} , & \snr_i \leq \gamma \leq \snr_{i+1} \\
0, & \snr_K < \gamma
\end{array}. \right.
\end{eqnarray}
\end{thm}
\begin{IEEEproof}
An alternative proof to the one given in \cite[section V.C]{StatisticalPhysics} is given in the Appendix.
\end{IEEEproof}
An example of a two-layered optimal Gaussian superposition code is depicted in Figure \ref{fig:1}, and a 4-layered optimal Gaussian superposition code is depicted in Figure \ref{fig:2}.

\begin{figure}
\begin{center}
\setlength{\unitlength}{.1cm}
    \includegraphics[width=0.45\textwidth]{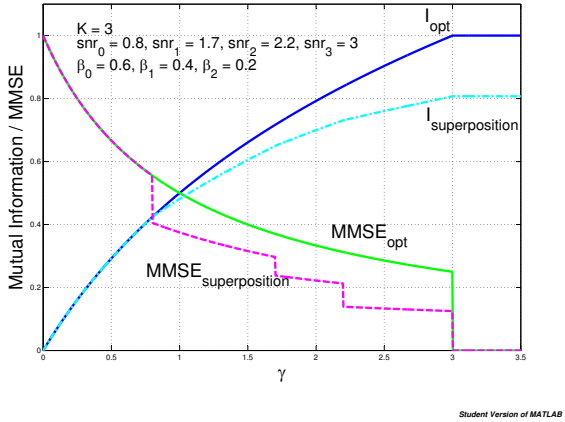}
    \caption{The mutual information and $\MSEcode(\gamma)$ of a 4-layers superposition code with $(\snr_0, \snr_1, \snr_2, \snr_3) = (0.8,1.7,2.2,3)$ and $(\beta_0, \beta_1, \beta_2) = (0.6,0.4, 0.3)$ and the mutual information and $\MSEcode(\gamma)$ of an optimal code for rate $\snr_3$.
    } \label{fig:2}
\end{center}
\end{figure}

\section{Problem Formulation} \label{sec:problem}
As stated, we are examining the scalar additive Gaussian channel, through which we transmit length $\dim$ codewords. For this setting we investigate the trade-off between rate and MMSE. This trade-off can be formalized in two equivalent manners. The first:

Assuming a pair of SNR points $(\snr_0, \snr_1)$ where $\snr_0 < \snr_1$, what is the solution of the following optimization problem:
\begin{align} \label{eq:singleConstraintOptimization1}
\max \quad & \I( \snr_1 ) \nonumber \\
\rm{s.t.} \quad & \MSEcode(\snr_0) \leq \frac{\beta}{1 + \beta \snr_0}
\end{align}
for some $\beta \in [0,1]$.

Alternatively, an equivalent form of the above problem is:
\begin{align} \label{eq:singleConstraintOptimization2}
\min \quad & \MSEcode(\snr_0) \nonumber \\
\rm{s.t.} \quad & \I( \snr_1 ) \geq \frac{1}{2} \log \left( 1 + \alpha \snr_1 \right)
\end{align}
for some $\alpha \in [0,1]$.
The exact connection between the two optimization problems, and the parameters $\beta$ and $\alpha$, will be made clear in Section \ref{sec:singleMMSEConstraint}.
The problem can also be extended to the general $K$ MMSE constraints as follows:

Assume a $K+1$ set of SNR points $(\snr_0, \snr_1, \cdots, \snr_K)$ such that $\snr_0 < \snr_1 < \cdots < \snr_K$ ($K \geq 1$ is some natural number). What is the solution of the following optimization problem:
\begin{align} 
\max \quad & \I( \snr_K ) \nonumber \\
\rm{s.t.} \quad & \MSEcode(\snr_i) \leq \frac{\beta_i}{1 + \beta_i \snr_i}, \quad \forall i \in \{ 0,1,\ldots, K-1\} \nonumber
\end{align}
for some $\beta_i \in [0,1]$, $i \in \{0,1,\ldots, K-1\}$, such that
\begin{align}
\sum_{i=0}^{K-1} \beta_i \leq 1 \quad \rm{and} \nonumber \\
\beta_{K-1} < \beta_{K-2} < \cdots < \beta_1 < \beta_0 \nonumber
\end{align}


\section{Single MMSE Constraint} \label{sec:singleMMSEConstraint}
In this section we present the main result of this paper, answering the following question: what is a maximum possible rate given a specific MMSE constraint at some lower SNR? In other words, we provide a solution to the optimization problem given in (\ref{eq:singleConstraintOptimization1}) (or alternatively, (\ref{eq:singleConstraintOptimization2})). We first give the main results and then detail the proofs in the subsequent subsections.

\subsection{Main Results} \label{ssec:singleConstraintMainResults}
The main result is given in the next theorem.
\begin{thm} \label{thm:singleConstraint}
Assume a pair of SNRs, $(\snr_0, \snr_1)$ such that $\snr_0 < \snr_1$. The solution of the following optimization problem,
\begin{align} \label{eq:thm:singleConstraintMainResult}
\max \quad & \I( \snr_1 ) \nonumber \\
\rm{s.t.} \quad & \MSEcode(\snr_0) \leq \frac{\beta}{1 + \beta \snr_0}
\end{align}
for some $\beta \in [0,1]$, is the following
\begin{align} \label{eq:superpositionRate}
\I(\snr_1) = \frac{1}{2} \log \left( 1 + \beta \snr_1 \right) + \frac{1}{2} \log \left( \frac{1 + \snr_0}{1 + \beta \snr_0} \right)
\end{align}
and is attainable when using the two-layered optimal Gaussian superposition codebook designed for $(\snr_0, \snr_1)$ with a rate-splitting coefficient $\beta$.
\end{thm}
The proof of this theorem is given in subsection \ref{ssec:proofTheoremSingleConstraint}.

An interesting question to ask is whether there could be a different code that can attain maximum rate under the MMSE constraint at $\snr_0$ (\ref{eq:thm:singleConstraintMainResult}) and also provide better MMSE for other values of SNR. The answer is to the negative, and is given in the next theorem.
\begin{thm} \label{thm:superpositionOptimality}
From the set of reliable codes of rate $\rate_c = \frac{1}{2} \log \left( 1 + \beta \snr_1 \right) + \frac{1}{2} \log \left( \frac{1 + \snr_0}{1 + \beta \snr_0} \right)$, complying with the MMSE constraint at $\snr_0$, the two-layered optimal Gaussian superposition codebook designed for $(\snr_0, \snr_1)$ with a rate-splitting coefficient $\beta$, provides the minimum MMSE for all SNRs.
\end{thm}

\subsection{Proof of Theorem \ref{thm:singleConstraint}} \label{ssec:proofTheoremSingleConstraint}
\begin{IEEEproof}
It is simple to verify that the two-layered optimal Gaussian superposition codebook designed for $(\snr_0, \snr_1)$ with a rate-splitting coefficient $\beta$, complies with the above MMSE constraint and attains the maximum rate. Thus, the focus of the remainder of the proof is on deriving a tight upper bound on the rate. We first prove the equivalent optimization problem, depicted in (\ref{eq:singleConstraintOptimization2}), and derive a lower bound on $\MSEcode(\snr_0)$ given a code, designed for reliable transmission at $\snr_1$ of rate $\rate_c = \frac{1}{2} \log \left(1 + \alpha \snr_1  \right)$.

If $ \alpha \snr_1 \leq \snr_0 \leq 1$ the lower bound is trivially zero using the optimal Gaussian codebook designed for $\alpha \snr_1$.
Thus, we assume $\snr_0 < \alpha \snr_1$.

Using the trivial upper bound on $\I(\snr_0) \leq \frac{1}{2} \log \left(1 + \snr_0 \right)$ (due to maximum entropy), we can lower bound the following difference, for any $\snr_0 < \alpha \snr_1$:
\begin{eqnarray} \label{eq:trivialUpperBound}
\I( \snr_1 ) - \I( \snr_0) \geq \I( \snr_1 ) - \frac{1}{2} \log \left( 1 + \snr_0  \right).
\end{eqnarray}
Using the I-MMSE relationship (\ref{eq:IMMSErelationship3_infty}), the above translates to the following inequality:
\begin{align} \label{eq:IMMSEusage}
\frac{1}{2} \int_{\snr_0}^{\snr_1} \MSEcode(\gamma) \dd \gamma & \geq \rate_c - \frac{1}{2} \log \left( 1 +  \snr_0  \right) \\
& = \frac{1}{2} \log \left( 1 + \alpha \snr_1 \right) - \frac{1}{2} \log \left( 1 +  \snr_0  \right). \nonumber
\end{align}
Defining $\d$ through the following equality:
\begin{multline}
\frac{1}{2} \log \left( 1 + \alpha \snr_1  \right) - \frac{1}{2} \log \left( 1 +  \snr_0  \right) =  \\
\frac{1}{2} \log \left( 1 + \d \snr_1  \right) - \frac{1}{2} \log \left( 1 + \d \snr_0 \right) .
\end{multline}
it is simple to check that for $\snr_0 < \alpha \snr_1$, $\d$ is in the range of $(0,1)$. Now we can continue with equation (\ref{eq:IMMSEusage}):
\begin{align} \label{eq:beforeSingleCrossing}
\frac{1}{2} \int_{\snr_0}^{\snr_1} \MSEcode(\gamma) \dd \gamma & \geq \frac{1}{2} \log \left( 1 + \d \snr_1  \right) - \frac{1}{2} \log \left( 1 + \d \snr_0  \right) \nonumber \\
& = \frac{1}{2} \int_{\snr_0}^{\snr_1} \mmse_G (\gamma) \dd \gamma.
\end{align}
where $\mmse_G(\gamma)$ is the MMSE assuming a Gaussian random variable with variance $\d$ transmitted through the additive Gaussian channel at SNR equal to $\gamma$. The single crossing point property (Theorem \ref{thm:ScalarUniqueCrossingPoint}) tells us that $\MSEcode(\gamma)$ and $\mmse_G (\gamma)$ cross each other at most once, and after that crossing point $\mmse_G (\gamma)$ remains an upper bound.
From the inequality in (\ref{eq:beforeSingleCrossing}) we can thus conclude that the single crossing point, if exists, must occur in the region $(\snr_0, \infty)$.
Thus, for $\snr_0$ we have the following lower bound:
\begin{align} \label{eq:MMSElowerBoundFinalStep}
\MSEcode(\snr_0) & \geq \frac{\d(\snr_0)}{1 + \d(\snr_0) \snr_0} = \frac{\alpha \snr_1 - \snr_0}{\snr_1 - \snr_0} \frac{1}{1 + \snr_0}
\end{align}
Note that $\d(\cdot)$ is a function of $\snr_0$.

In terms of the equivalent optimization problem, given in equation (\ref{eq:singleConstraintOptimization1}), the case of $\alpha \snr_1 \leq  \snr_0$ is equivalent to a zero constraint on the MMSE, that is, $\beta = 0$. For $\beta \in (0,1]$ the lower bound derived in (\ref{eq:MMSElowerBoundFinalStep}) can be written in terms of the constraint on MMSE, resulting with the following connection between the two parameters:
\begin{align} \label{eq:connectingParameters}
\alpha = \frac{ \beta (\snr_1 - \snr_0) + \snr_0 ( 1 + \beta \snr_1)}{\snr_1 (1 + \beta \snr_0)}.
\end{align}
Substituting this connection in $\rate_c = \frac{1}{2} \log \left(1 + \alpha \snr_1  \right)$ results with the superposition rate given in (\ref{eq:superpositionRate}).
\end{IEEEproof}

\subsection{Proof of Theorem \ref{thm:superpositionOptimality}} \label{ssec:proofTheoremSuperpositionOptimality}
\begin{IEEEproof}
The code complies with the following constraint:
\begin{align} \label{eq:IMMSEmainUsgae2Single}
\MSEcode(\snr_0) \leq \mmse_{G}(\snr_0) = \frac{\beta}{1 + \beta \snr_0}
\end{align}
where $\mmse_{G}(\snr_0)$ denotes the MMSE of the estimation of a Gaussian random variable, $X_{G}$, with zero mean and variance $\beta$, from $Y = \sqrt{\snr_0} X_{G} + N$, where $N \sim \mathcal{N}(0,1)$. Thus,
\begin{align}
\q(\rvec{x}, \beta, \snr_0) = \mmse_{G}(\snr_0) - \MSEcode(\snr_0) \geq 0.
\end{align}
According to Theorem \ref{thm:ScalarUniqueCrossingPoint} the function $\q(\rvec{x}, \beta, \gamma)$ has no nonnegative-to-negative zero crossings, thus we may conclude that,
\begin{align} \label{eq:IMMSEmainUsgaeSingle}
\q(\rvec{x}, \beta, \gamma) \geq 0 \Leftrightarrow \MSEcode(\gamma) \leq \mmse_{G}(\gamma) \quad \forall \gamma \geq \snr_0
\end{align}
and derive the following upper bound,
\begin{align}
\I( \snr_{1} ) - \I( \snr_0 ) & = \frac{1}{2} \int_{\snr_0}^{\snr_{1}} \MSEcode(\gamma) \d \gamma \nonumber \\
& \leq \frac{1}{2} \int_{\snr_0}^{\snr_{1}} \mmse_{G} (\gamma) \d \gamma \nonumber \\
& = \frac{1}{2} \log \left( \frac{1 + \beta \snr_{1}}{1 + \beta \snr_0} \right).
\end{align}
On the other hand, since we are assuming a code that attains the maximum rate we can lower bound the above difference using the maximum entropy theorem,
\begin{align}
\I( \snr_{1} ) - \I( \snr_0 ) & \geq \rate_c - \frac{1}{2} \log \left( 1 + \snr_0 \right) \nonumber \\
& = \frac{1}{2} \log \left( \frac{1 + \beta \snr_{1}}{1 + \beta \snr_0} \right).
\end{align}
Thus, we conclude that any code complying with the MMSE constraint and obtaining the maximum rate obtains the above two inequalities with equality. In order to attain the upper bound,
\begin{align}
\frac{1}{2} \int_{\snr_0}^{\snr_{1}} \MSEcode(\gamma) \d \gamma = \frac{1}{2} \int_{\snr_0}^{\snr_{1}} \mmse_{G} (\gamma) \d \gamma \nonumber
\end{align}
however, due to (\ref{eq:IMMSEmainUsgaeSingle}) we have,
\begin{align}
\MSEcode(\gamma) = \mmse_{G} (\gamma), \quad \forall \gamma \in [\snr_0, \snr_1] \nonumber
\end{align}
In order to attain the lower bound, given that $\I( \snr_{1} ) = \rate_c$, we require,
\begin{align}
\I( \snr_0 ) = \frac{1}{2} \log \left( 1 + \snr_0 \right) \nonumber
\end{align}
which guarantees $ \MSEcode(\gamma) = \frac{1}{2} \log \left( 1 + \gamma \right)$ for all $\gamma \in [0, \snr_0]$. Finally, for $\gamma \in [\snr_1, \infty)$, since we assume codebooks that are reliably decoded at $\snr_1$,  $\MSEcode(\gamma) = 0$. To conclude, we have shown that for any code complying with the MMSE constraint and attaining the maximum rate, the $\MSEcode(\gamma)$ function is defined for all $\gamma \in [0, \infty)$, and thus also the mutual information. This concludes our proof.
\end{IEEEproof}

\section{Multi-MMSE Constraints} \label{sec:multiMMSEConstraints}
In this section we extend the results for the single MMSE constraint, given in the previous section, to $K$ MMSE constraints, and examine the same question: under these $K$ MMSE constraints, what is the maximum possible rate?

\subsection{Main Results}
The main result of this section is given in the next theorem.
\begin{thm} \label{thm:Kextension}
Assume a set of SNRs, $(\snr_0, \snr_1, \cdots, \snr_K)$ such that $\snr_0 < \snr_1 < \cdots < \snr_K$ ($K \geq 1$ is some natural number). The solution of the following optimization problem,
\begin{align}
\max \quad & \I( \snr_K ) \nonumber \\
\rm{s.t.} \quad & \MSEcode(\snr_i) \leq \frac{\beta_i}{1 + \beta_i \snr_i}, \quad \forall i \in \{ 0,1,\ldots, K-1\} \nonumber
\end{align}
for some $\beta_i \in [0,1]$, $i \in \{0,1,\ldots, K-1\}$, such that
\begin{align}
\sum_{i=0}^{K-1} \beta_i \leq 1 \quad \rm{and} \nonumber \\
\beta_{K-1} < \beta_{K-2} < \cdots < \beta_1 < \beta_0 \nonumber
\end{align}
is the following
\begin{multline}
\I(\snr_K) = \frac{1}{2} \log \left( \frac{1 + \snr_0}{1 + \beta_0 \snr_0} \prod_{j=1}^{K-1} \frac{ 1 + \beta_{j-1} \snr_j}{1 + \beta_j \snr_j} \right) + \\ \frac{1}{2} \log \left( 1 + \beta_{K-1} \snr_K \right)
\end{multline}
and is attainable when using the optimal $K$-layers Gaussian superposition codebook designed for $(\snr_0, \snr_1, \cdots, \snr_K)$ with rate-splitting coefficients $(\beta_0, \cdots, \beta_{K-1})$.

Additional constraints of the following form:
\begin{align}
\MSEcode(\snr_{\ell}) \leq \frac{\beta_{\ell}}{1 + \beta_{\ell} \snr_{\ell}}
\end{align}
for $\snr_{i-1} \leq \snr_{\ell} \leq \snr_{i}$ when $\beta_{\ell} \geq \beta_{i-1}$, do not affect the above result.
\end{thm}

Theorem \ref{thm:Kextension} states that $K$-layers superposition codes attain the maximum possible rate at $\snr_K$ under a set of $K$ MMSE constraints at lower SNRs. However, there might be a different codebook with this property, which also has some other desirable properties. In the next theorem we prove that the behavior of the MMSE and the mutual information as a function of the $\snr$, for any code attaining the maximum rate under the set of MMSE constraints, is known for all $\snr$, and are those of $K$-layers superposition codes. Thus, no other code can outperform superposition codes in this sense.

\begin{thm} \label{thm:KsuperpositionOptimal}
The $\MSEcode(\gamma)$ (and thus also $\I(\gamma)$) of any code attaining the maximum rate at $\snr_K$, under the MMSE constraints, defined in Theorem \ref{thm:Kextension}, is known for all $0 \leq \gamma$, and is that of the $K$-layers superposition codebook.
\end{thm}

\subsection{Proof of Theorem \ref{thm:Kextension}} \label{ssec:proofTheoremKextension}
\begin{IEEEproof}
\begin{figure*}[!t]
\normalsize
\setcounter{MYtempeqncnt}{\value{equation}}
\setcounter{equation}{37}
\begin{align} \label{eq:exactDerivation}
\I(\snr_K) 
& \leq \frac{1}{2} \log \left( 1 + \beta_0 \snr_1 \right) + \frac{1}{2} \log \left( \frac{1 + \snr_0}{1 + \beta_0 \snr_0} \right) + \sum_{i =1}^{K-1} \frac{1}{2} \log \left( \frac{1 + \beta_i \snr_{i+1}}{1 + \beta_i \snr_i} \right) \nonumber \\
& = \frac{1}{2} \log \left( \frac{1 + \snr_0}{1 + \beta_0 \snr_0} \right) + \frac{1}{2} \log \left( (1 + \beta_0 \snr_1) \frac{1 + \beta_1 \snr_{2}}{1 + \beta_1 \snr_1} \frac{1 + \beta_2 \snr_{3}}{1 + \beta_2 \snr_2} \frac{1 + \beta_3 \snr_{4}}{1 + \beta_3 \snr_3} \cdots \frac{1 + \beta_{K-2} \snr_{{K-1}}}{1 + \beta_{K-2} \snr_{K-2}} \frac{1 + \beta_{K-1} \snr_{{K}}}{1 + \beta_{K-1} \snr_{K-1}} \right) \nonumber \\
& = \frac{1}{2} \log \left( \frac{1 + \snr_0}{1 + \beta_0 \snr_0} \right) + \frac{1}{2} \log \left(  \frac{1 + \beta_0 \snr_1}{1 + \beta_1 \snr_1} \frac{1 + \beta_1 \snr_{2}}{1 + \beta_2 \snr_2} \frac{1 + \beta_2 \snr_{3}}{1 + \beta_3 \snr_3} \cdots \frac{1 + \beta_{K-2} \snr_{K-1}}{1 + \beta_{K-1} \snr_{K-1}}  \right) + \frac{1}{2} \log \left( {1 + \beta_{K-1} \snr_{{K}}} \right) \nonumber \\
& = \frac{1}{2} \log \left(  \frac{1 + \snr_0}{1 + \beta_0 \snr_0} \prod_{j=1}^{K-1} \frac{1 + \beta_{j-1} \snr_j}{1 + \beta_j \snr_j} \right) + \frac{1}{2} \log \left( {1 + \beta_{K-1} \snr_{{K}}} \right)
\end{align}
\setcounter{equation}{\value{MYtempeqncnt}}
\hrulefill
\vspace*{4pt}
\end{figure*}
It is simple to verify that the optimal Gaussian $K$-layers superposition codebook (Theorem \ref{thm:superpositionMI_MMSE}) complies with the above MMSE constraints and attains the maximum rate. Thus, we need to derive a tight upper bound on the rate.
Deriving the upper bound begins with the usage of Theorem \ref{thm:singleConstraint}. Due to the constraint at $\snr_0$:
\begin{align}
\MSEcode(\snr_0) \leq \frac{\beta_0}{1 + \beta_0 \snr_0}
\end{align}
we have the following upper bound
\begin{align} \label{eq:ineuqliaty1}
\I(\snr_1) \leq \frac{1}{2} \log \left( 1 + \beta_0 \snr_1 \right) + \frac{1}{2} \log \left( \frac{1 + \snr_0}{1 + \beta_0 \snr_0} \right) .
\end{align}
The other constraints, for $i \in \{1,2, \ldots, K-1\}$, can be written as follows,
\begin{align}
\MSEcode(\snr_i) \leq \frac{\beta_i}{1 + \beta_i \snr_i} = \mmse_{G_i}(\snr_i)
\end{align}
where $\mmse_{G_i}(\snr_i)$ denotes the MMSE of the estimation of a Gaussian random variable, $X_{G_i}$, with zero mean and variance $\beta_i$, from $Y = \sqrt{\snr_i} X_{G_i} + N$, where $N \sim \mathcal{N}(0,1)$. Thus,
\begin{align}
\q(\rvec{x}, \beta_i, \snr_i) = \mmse_{G_i}(\snr_i) - \MSEcode(\snr_i) \geq 0.
\end{align}
According to Theorem \ref{thm:ScalarUniqueCrossingPoint} the function $\q(\rvec{x}, \beta_i, \gamma)$ has no nonnegative-to-negative zero crossings, thus we may conclude that,
\begin{align} \label{eq:IMMSEmainUsgae}
\q(\rvec{x}, \beta_i, \gamma) \geq 0 \Leftrightarrow \MSEcode(\gamma) \leq \mmse_{G_i}(\gamma) \quad \forall \gamma \geq \snr_i.
\end{align}
This allows us to provide a tight upper bound on the following difference:
\begin{align} \label{eq:ineuqliaty2}
\I( \snr_{i+1} ) - \I( \snr_i ) & = \frac{1}{2} \int_{\snr_i}^{\snr_{i+1}} \MSEcode(\gamma) \d \gamma \nonumber \\
& \leq \frac{1}{2} \int_{\snr_i}^{\snr_{i+1}} \mmse_{G_i} (\gamma) \d \gamma \nonumber \\
& = \frac{1}{2} \log \left( \frac{1 + \beta_i \snr_{i+1}}{1 + \beta_i \snr_i} \right).
\end{align}
Now, we can write the objective function as follows:
\begin{align} \label{eq:objFunc}
\I(\snr_K) = \I( \snr_1)  + \sum_{i =1}^{K-1} \left[ \I( \snr_{i+1}) - \I(\snr_i) \right]
\end{align}
Using (\ref{eq:ineuqliaty1}) and (\ref{eq:ineuqliaty2}) we can bound (\ref{eq:objFunc}) as shown in (\ref{eq:exactDerivation}) at the top of the next page.
\addtocounter{equation}{1}

Now, according to (\ref{eq:IMMSEmainUsgae}) we have that any additional constraint,
$\MSEcode(\snr_{\ell}) \leq \frac{\beta_{\ell}}{1 + \beta_{\ell} \snr_{\ell}}$
for $\snr_{i-1} \leq \snr_{\ell} \leq \snr_{i}$ when $\beta_{\ell} \geq \beta_{i-1}$, is already complied with, since
\begin{align}
\MSEcode(\snr_{\ell}) \leq \frac{\beta_{i-1}}{1 + \beta_{i-1} \snr_{\ell}} \leq \frac{\beta_{\ell}}{1 + \beta_{\ell} \snr_{\ell}}
\end{align}
and thus, does not affect the result.
This concludes our proof.
\end{IEEEproof}

\subsection{Proof of Theorem \ref{thm:KsuperpositionOptimal}} \label{ssec:proofTheoremKsuperpositionOptimal}
\begin{IEEEproof}
Due to the set of $K$ constraints and following the steps that lead to (\ref{eq:IMMSEmainUsgae}) in the proof of Theorem \ref{thm:Kextension} we can conclude that
\begin{align} \label{eq:IMMSEmainUsgae2}
\MSEcode(\gamma) & \leq \mmse_{G_i}(\gamma) = \frac{\beta_i}{1 + \beta_i \gamma}, \quad \forall \gamma \geq \snr_i
\end{align}
for $i \in \{0,1,2, \ldots, K-1\}$, where $\mmse_{G_i}(\snr_i)$ denotes the MMSE of the estimation of a Gaussian random variable, $X_{G_i}$, with zero mean and variance $\beta_i$, from $Y = \sqrt{\snr_i} X_{G_i} + N$, where $N \sim \mathcal{N}(0,1)$.

In the proof of Theorem \ref{thm:Kextension}, equation (\ref{eq:ineuqliaty2}), we have seen that the above property can be used to construct the following upper bounds
\begin{align} \label{eq:ineuqliaty3}
\I( \snr_{i+1} ) - \I( \snr_i ) & \leq \frac{1}{2} \int_{\snr_i}^{\snr_{i+1}} \mmse_{G_i} (\gamma) \d \gamma \nonumber \\
& = \frac{1}{2} \log \left( \frac{1 + \beta_i \snr_{i+1}}{1 + \beta_i \snr_i} \right).
\end{align}
From these upper bounds we can obtain the following
\begin{align} \label{eq:upperBound}
\I( \snr_{K} ) - \I( \snr_0 ) &  = \frac{1}{2} \int_{\snr_0}^{\snr_{K}} \MSEcode(\gamma) \d \gamma \nonumber \\
& = \sum_{i = 0}^{K-1} \frac{1}{2} \int_{\snr_i}^{\snr_{i+1}} \MSEcode(\gamma) \d \gamma \nonumber \\
& \leq \sum_{i = 0}^{K-1} \frac{1}{2} \int_{\snr_i}^{\snr_{i+1}} \mmse_{G_i} (\gamma) \d \gamma \nonumber \\
& = \sum_{i = 0}^{K-1} \frac{1}{2} \log \left( \frac{1 + \beta_i \snr_{i+1}}{1 + \beta_i \snr_i} \right) \nonumber \\
& = \frac{1}{2} \log \prod_{i=0}^{K-1} \left(  \frac{1 + \beta_i \snr_{i+1}}{1 + \beta_i \snr_i}  \right).
\end{align}
On the other hand, we can lower bound the above difference:
\begin{align} \label{eq:lowerBound}
\I( \snr_{K} ) - \I( \snr_0 ) & \geq \rate_c - \frac{1}{2} \log \left( 1 + \snr_0 \right) \nonumber \\
& = \frac{1}{2} \log \prod_{i=0}^{K-1} \left(  \frac{1 + \beta_i \snr_{i+1}}{1 + \beta_i \snr_i}  \right)
\end{align}
where we used both the assumption that the code attains the maximum rate at $\snr_K$, under the MMSE constraints (Theorem \ref{thm:Kextension}), and the maximum entropy theorem to obtain the maximum mutual information at $\snr_0$. From (\ref{eq:upperBound}) and (\ref{eq:lowerBound}) we have
\begin{align}
\I( \snr_{K} ) - \I( \snr_0 ) = \frac{1}{2} \log \prod_{i=0}^{K-1} \left(  \frac{1 + \beta_i \snr_{i+1}}{1 + \beta_i \snr_i}  \right)
\end{align}
for any code attaining the maximum rate at $\snr_K$ under the MMSE constraints, given in Theorem \ref{thm:Kextension}. Looking at the upper bound (\ref{eq:upperBound}), this equality can be attained only if
\begin{multline}
\frac{1}{2} \int_{\snr_i}^{\snr_{i+1}} \MSEcode(\gamma) \d \gamma = \frac{1}{2} \int_{\snr_i}^{\snr_{i+1}} \mmse_{G_i} (\gamma) \d \gamma, 
\end{multline}
for all $i \in \{0,1,\ldots, K-1\}$.
Due to (\ref{eq:IMMSEmainUsgae2}) this is equivalent to
$\MSEcode(\gamma)= \mmse_{G_i}(\gamma) = \frac{\beta_i}{1 + \beta_i \gamma}$ 
for all $\snr_i \leq \gamma < \snr_{i+1}$.
Thus, we defined the function $\MSEcode(\gamma)$ for all $\gamma \in [\snr_0, \snr_K]$. Surely since this is a reliable code designed for $\snr_K$, we also have that $\MSEcode(\gamma) = 0$ for all $\gamma \geq \snr_K$. The only region that remains to be determined is $\gamma \in [0, \snr_0]$. Since the lower bound, (\ref{eq:lowerBound}), is attained with equality and $\I( \snr_{K} ) = \rate_c$
we have
\begin{align}
\I( \snr_0 ) = \frac{1}{2} \log \left( 1 + \snr_0 \right)
\end{align}
which guarantees that $\MSEcode(\gamma) = \frac{1}{1 + \gamma}$ for all $\gamma \in [0, \snr_0]$. This concludes our proof.
\end{IEEEproof}

\section{Finite Length Code} \label{sec:finiteLength}
We now extend the single MMSE constraint result, given in section \ref{sec:singleMMSEConstraint}, to the case of finite length codes. In this case the code is not fully reliable, but rather has a small probability of error, denoted as $\pdffun_e$. In the case that this error probability in unknown precisely, one may upper bound it using basic properties of the code \cite{SasonShamaiFnt}.
\begin{cor} \label{cor:MMSELowerBoundsFiniteLengthCode}
Assume a finite length code of rate $\rate_c = \frac{1}{2} \log \left(1 + \alpha \snr_1  \right)$, designed for transmission at $\snr_1$ with error probability $\pdffun_e$.
For any $\snr_0 < \alpha \snr_1$ we have the following lower bound,
\begin{multline}
\MSEcoden(\snr_0) \geq \\  \frac{1 + \alpha \snr_1 - (1 +\snr_0) 2^{\Hb{ \pdffun_e}{2} } (1 + \alpha \snr_1 )^{\pdffun_e} }{2^{ \Hb{ \pdffun_e}{2}} (1 + \alpha \snr_1 )^{\pdffun_e} \left[ \snr_1 - \snr_0 + \snr_0(\snr_1 - \snr_0) \right]}
\end{multline}
where $\ss{h}_b \left( \cdot  \right)$ stands for the binary entropy function.
\end{cor}
\begin{IEEEproof}
Due to Fano's inequality \cite{CoverThomas} we have,
\begin{align}
\I( \snr_1 ) & = \rate_c - \HdC{ \rvec{x}}{\rvec{y}(\snr_1)} \nonumber \\
& \geq \rate_c - \Hb{ \pdffun_e}{1} - \frac{1}{\dim} \pdffun_e \log \left( 2^{\dim \rate_c} - 1 \right) \nonumber \\
& \geq \frac{1}{2} \log \left( 1 + \alpha \snr_1 \right) - \frac{1}{2} \log 2^{\Hb{ \pdffun_e}{2} } - \frac{1}{2} \log 2^{2 \pdffun_e \rate_c} \nonumber \\
& = \frac{1}{2} \log \left( 1 + \alpha \snr_1 \right) - \frac{1}{2} \log 2^{ \Hb{ \pdffun_e}{2}} - \frac{1}{2} \log  \left( 1 + \alpha \snr_1 \right)^{\pdffun_e} \nonumber \\
& = \frac{1}{2} \log \left[ \left( 1 + \alpha \snr_1 \right)^{1 - \pdffun_e}  2^{-\Hb{ \pdffun_e}{2} } \right].
\end{align}
Now, using this lower bound in (\ref{eq:IMMSEusage}) we obtain a new value for the parameter $\d$,
\begin{align} 
\d = \frac{1 + \alpha \snr_1 - (1 + \snr_0)2^{\Hb{ \pdffun_e}{2}}\left( 1 + \alpha \snr_1 \right)^{\pdffun_e} }{2^{\Hb{ \pdffun_e}{2}}\left( 1 + \alpha \snr_1 \right)^{\pdffun_e} \snr_1 (1 + \snr_0)- \snr_0( 1 + \alpha \snr_1)}. \nonumber
\end{align}
Placing the above in the lower bound of (\ref{eq:MMSElowerBoundFinalStep}) we obtain the desired result. This concludes our proof.
\end{IEEEproof}

\begin{rem} \label{rem:finitLengthCode}
Note that contrary to the case of $\dim \to \infty$, since the code is not fully reliable, we do not have that $\MSEcoden(\gamma) = 0$ for all $\gamma \geq \snr_1$. Furthermore, we do not have a trivial lower bound of zero for $\gamma \geq \alpha \snr_1$.
\end{rem}

As an example for the above lower bound we can examine regular LDPC codes, for which the tangential-sphere bound (TSB) provides a good upper bound on $\pdffun_e$ \cite{SasonShamaiFnt}. Using the results of \cite[pp. 78]{SasonShamaiFnt}, we have that a regular (6, 12)-LDPC code of block length $\dim = 5 K$ and rate $\rate_c = 0.5$, obtains $\pdffun_e = 10^{-5}$ at
$\snr_1 = 2.5179$ and $\alpha \snr_1 = 1$. The lower bound of Corollary \ref{cor:MMSELowerBoundsFiniteLengthCode}, for $\gamma <  \alpha \snr_1 $, is given in Figure \ref{fig:2} (in blue), together with the uncoded MMSE \cite[eq. (17)]{IMMSE}, which provides an upper bound (in red). Note that for ``bad'' LDPC codes, tighter upper bounds can be provided using Belief-Propagation analysis (or the I-MMSE approach) \cite{Bennatan2}. However, these upper bounds improve the upper bound for SNRs nearing $\snr_1$ (for which the lower bound of Corollary \ref{cor:MMSELowerBoundsFiniteLengthCode} is useless) and, on the other hand, for low SNRs consolidate with the upper bound (depicted in red in Figure \ref{fig:2}) \cite{Bennatan2}.

\begin{figure}
\begin{center}
\setlength{\unitlength}{.1cm}
    \includegraphics[width=0.55\textwidth]{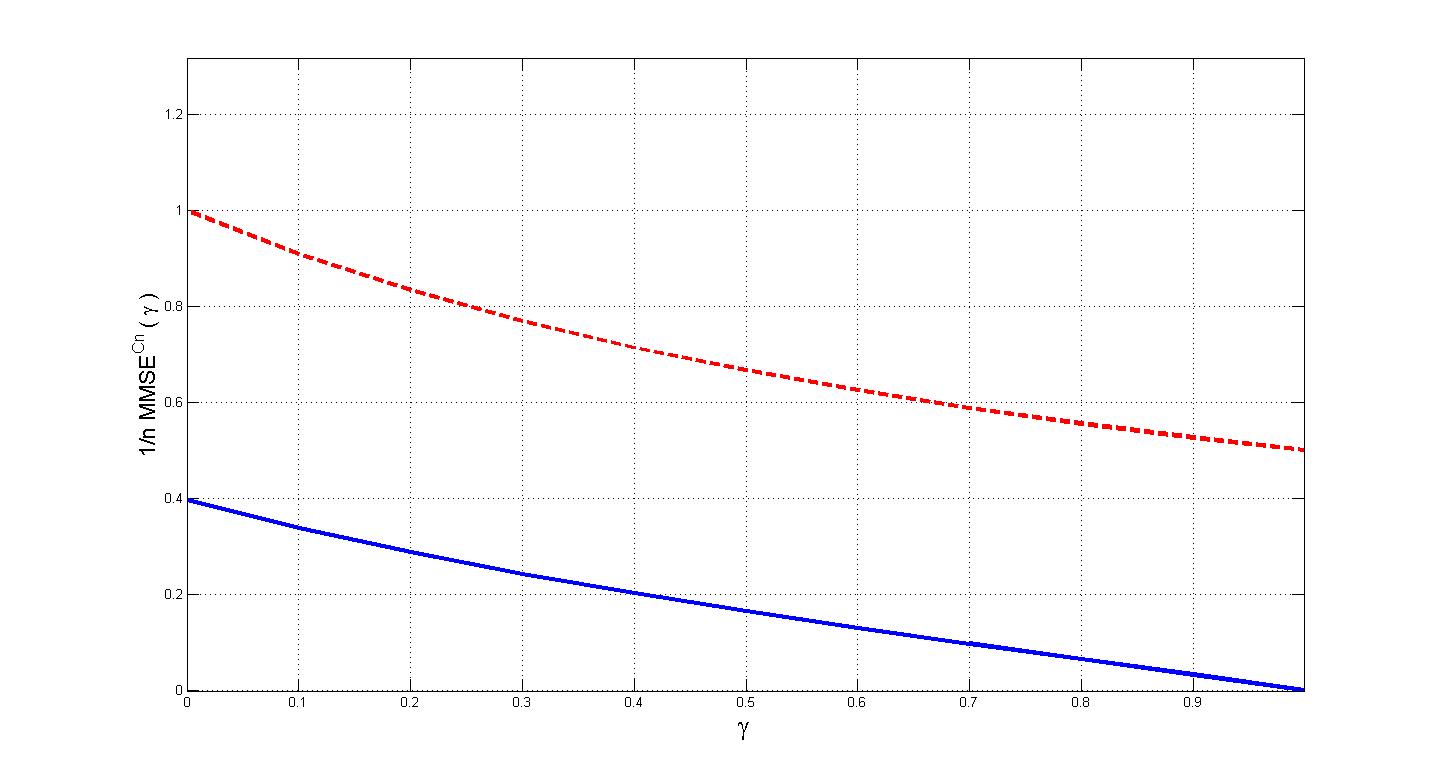}
    \caption{The lower bound on the $\MSEcoden$ of a regular (6,12)-LDPC code of length $\dim = 5K$, $\rate_c = 0.5$, and $\pdffun_e = 10^{-5}$ at $\snr_1 = 2.5179$ (data taken from \cite[pp. 78]{SasonShamaiFnt}), given for $0 < \gamma < \alpha \snr_1 = 1$ (in solid). The uncoded MMSE is given in dashed.
    } \label{fig:2}
\end{center}
\end{figure}

\section{The Mutual Information ``Disturbance'' Measure} \label{sec:mutualInformationDisturbance}
Bandemer and El Gamal \cite{Bandemer} suggested a difference measure of ``disturbance'' to a receiver not interested in the transmitted message. Bandemer and El Gamal examined discrete memoryless channels and derived a single-letter expression for the problem with a single ``disturbance'' constraint. Applying the single-letter to the scalar Gaussian case they obtain the following result,
\begin{cor}[\cite{Bandemer}] \label{cor:BandemerSingleConstraint}
The rate-disturbance region of the Gaussian channel for the pair of SNRs $(\snr_0, \snr_1)$ is
\begin{align}
\rate & \leq  \frac{1}{2} \log \left( 1 + \alpha \snr_1 \right) \nonumber \\
\rate_d & \geq \frac{1}{2} \log \left( 1 + \alpha \snr_0 \right)
\end{align}
for any $\alpha \in [0,1]$. The maximum rate is attained by an optimal Gaussian codebook designed for $\snr_1$ with limited power of $\alpha$.
\end{cor}
\begin{IEEEproof}
The above result, which originally has been proved by the entropy power inequality \cite{Bandemer}, can also be derived directly from the I-MMSE formulation. Starting from the disturbance rate, since
\begin{align}
0 \leq \I_{\dim}(\snr_0) \leq \frac{1}{2} \log \left(1 + \snr_0 \right)
\end{align}
there exists an $\alpha \in [0,1]$ such that,
\begin{align}
\I_{\dim}(\snr_0) = \frac{1}{2} \log \left(1 + \alpha \snr_0 \right) .
\end{align}
Using the I-MMSE approach, the above can be written as follows,
\begin{align}
\frac{1}{2} \int_{0}^{\snr_0} \MSEcoden(\gamma) \d \gamma = \frac{1}{2} \int_{0}^{\snr_0} \mmse_G(\gamma) \d \gamma.  \nonumber
\end{align}
According to Theorem \ref{thm:ScalarUniqueCrossingPoint} we conclude that $\MSEcode(\gamma)$ and $\mmse_G(\gamma)$ are either equal for all $\gamma$, or alternatively, cross each other once in the region $[0, \snr_0)$. In both cases we have,
\begin{align} \label{eq:BandemerScalarSingleConstraint}
\MSEcoden(\gamma) \leq \mmse_G(\gamma), \quad \forall \gamma \in [\snr_0, \infty) .
\end{align}
Now, upper bounding the rate,
\begin{align}
\I_{\dim}(\snr_1) & = \frac{1}{2} \log \left(1 + \alpha \snr_0 \right)  + \int_{\snr_0}^{\snr_1} \MSEcoden(\gamma) d \gamma \nonumber \\
& \leq \frac{1}{2} \log \left(1 + \alpha \snr_1 \right)
\end{align}
This concludes the I-MMSE based proof.
\end{IEEEproof}
Extending Corollary \ref{cor:BandemerSingleConstraint} to $K$ mutual information disturbance constraints, in the Gaussian channel, is trivial since only one of the constraints remains effective. The result is given in the next corollary.
\begin{cor} \label{cor:BandemerMultiConstraints}
Assume a set of SNRs, $(\snr_0, \snr_1,$ $\cdots, \snr_K)$, such that $\snr_0 < \snr_1 < \cdots < \snr_K$. The solution of
\begin{align}
\max \quad & \I_{\dim}( \snr_K ) \nonumber \\
\rm{s.t.} \quad & \forall i \in \{0,\cdots, K-1\}, \quad \I_{\dim}( \snr_i ) \leq \frac{1}{2} \log \left( 1 +  \alpha_i \snr_i \right) \nonumber
\end{align}
for some values $\alpha_i \in [0,1]$, is the following
\begin{align}
\I_{\dim}(\snr_K) = \frac{1}{2} \log \left( 1 + \alpha_{\ell} \snr_K \right) \nonumber
\end{align}
where $\alpha_{\ell}$, $\ell \in \{1, \cdots, K-1\}$, is defined such that
\begin{align}
\forall i \in \{0, \cdots, K-1 \} \quad \frac{1}{2} \log \left( 1 +  \alpha_{\ell} \snr_i \right) \leq \frac{1}{2} \log \left( 1 +  \alpha_i \snr_i \right) \nonumber
\end{align}
The maximum rate is attained, for any $\dim$, by choosing the input to be Gaussian with \iid components of variance $\alpha_{\ell}$. For $\dim \to \infty$ equality is also attained by an optimal Gaussian codebook designed for $\snr_K$ with limited power of $\alpha_{\ell}$.
\end{cor}

\section{Conclusions and Discussion} \label{sec:conclusions}
In this work we quantify the advantage of ``bad'' point-to-point codes, in terms of MMSE. These codes, that do not attain capacity, are heavily used in multi-user wireless networks. We show that the maximum possible rate of an MMSE constrained code is the rate of the corresponding optimal Gaussian superposition codebook. We also show that the MMSE and mutual information behavior as a function of SNR of any code attaining the maximum rate under the MMSE constraint, is known for all SNR. The result are then extended to $K$ MMSE constraints. We also provide a lower bound on the MMSE of finite codes.

As stated in the Introduction, the single MMSE constraint result provide the engineering insight to the good performance of the HK superposition scheme on the two-user interference channel, as shown in \cite{ETW}. Our results, showing that the HK superposition scheme is optimal MMSE-wise suggest that one cannot construct better codes of the type defined in \cite{Bennatan1} that will beat HK through the use of estimation. Note that, as mentioned in \cite[section V]{Bennatan1}, the codes constructed there have an important complexity advantage over HK codes.

The HK scheme is efficient in the two-user interference channel and only a simple approach in the general $K$-user interference channel. In other words, the MMSE-wise optimality of this scheme for the $K$ MMSE constrained problem is not sufficient to guarantee an efficient coding scheme. The reason being that the $K$ MMSE constrained problem is a huge simplification of the interference channel, as only a single message is transmitted and creates interference to $K$ receivers, whereas in the $K$-user interference channel, each receiver suffers interference from all other $K-1$ receivers. As well known, the interference alignment approach obtains, for certain interference channel coefficients, better results, in terms of rate and degrees of freedom, as compared to the HK scheme in the $K$-user interference channel. It has been shown that I-MMSE considerations based on information and MMSE dimension are useful also in these kind of problems, see \cite{WuDoF}, \cite{TuninettiKusers} and references therein.

In the previous section we have shown that the different disturbance measure suggested by Bandemer and El Gamal \cite{Bandemer} does not suggest rate-splitting in the scalar Gaussian channel, but rather an optimal Gaussian codebook of reduced power. Moreover, the extension to $K$ constraints reduces to a single effective constraint and also suggests an optimal Gaussian codebook of reduced power. On the other hand, the results of Bandemer and El Gamal are valid for any finite $\dim$ as opposed to our results which are given only for $\dim \to \infty$. To conclude, the two measures of disturbance are conceptually different. Finally, Bandemer and El Gamal \cite{Bandemer} also extended their work to the MIMO Gaussian channel, where optimality does requires rate-splitting codebooks. One of our challenges is to extend the MMSE constrained problem to the MIMO Gaussian channel. Note also that the results of Bandemer and El Gamal for the Gaussian channel

The main challenge, that also has significant implications on the design of actual codes, is the extension of the results given above to the finite $\dim$ case. In other words, what is the maximum mutual information given a constraint on the MMSE of a finite length code. This optimization problem is also interesting for $\dim = 1$, where no code is considered. It was conjectured in \cite{ShlomosLetter} that for the $\dim = 1$ case, the optimizing, finite variance, random variable is discrete.

In this work we proved that under MMSE constraints at lower SNRs, the optimal code, when $\dim \to \infty$, attaining maximum rate is a superposition codebook. This raises another challenge: what is the maximum possible rate if we further limit the discussion to single structured codes (still at $\dim \to \infty$)? In other words, what is the solution of the given optimization problems if we add an additional constraint that the MMSE curve does not exhibit phase transitions, and is continuous until $\snr_K$?

\appendix

\subsection{Proof of Theorem \ref{thm:superpositionMI_MMSE}} \label{appendix:proofSuperposition}
\begin{IEEEproof} The proof given here is an elaboration of the last paragraph in \cite[section V.C, The Gaussian Broadcast Channel]{StatisticalPhysics}, which provides the optimal Gaussian BC codebook viewpoint. We prove only the expressions of the two-layered optimal Gaussian superposition codebook. The extension to the general $L$-layers ($L > 1$) is straightforward.

Using the definition of an optimal Gaussian superposition codebook given in Definition \ref{dfn:superposition}, we have a Markov chain, $(\rvec{u}, \rvec{v}) - \rvec{x} - \rvec{y}(\gamma)$, and the mutual information can be written as follows:
\begin{align} \label{appendix:proofMIsuperposition}
\I_{\dim}(\gamma) & = \frac{1}{\dim } \Igen{\rvec{x}}{\rvec{y}(\gamma) = \sqrt{\gamma} \rvec{x} + \rvec{n}} \nonumber \\
& = \frac{1}{\dim }  \Igen{\rvec{u}, \rvec{x}}{\rvec{y}(\gamma)} \nonumber \\ 
& = \frac{1}{\dim }  \Igen{\rvec{u}}{\rvec{y}(\gamma)} + \frac{1}{\dim }  \Icond{\rvec{x}}{\rvec{y}(\gamma)}{\rvec{u}} .
\end{align}
We want to derive the limit, as $\dim \to \infty$, of the above expression. As we are examining a two-layered optimal Gaussian superposition codebook we have a pair of relevant SNR points, $( \snr_0, \snr_1)$.
We begin by examining $\Igen{\rvec{u}}{\rvec{y}(\gamma)}$ at SNRs below $\snr_0$, for $\dim \to \infty$.
At these SNRs the private message acts as additive Gaussian noise, since otherwise one could take advantage of that and transmit the common message at a higher rate, contradicting the capacity of the scalar Gaussian BC. Thus, we have, for $\dim \to \infty$,
\begin{eqnarray} \label{appendix:eq:privateMessage}
\Igen{\rvec{u}}{\rvec{y}(\gamma)} = \Igen{\rvec{u}}{\sqrt{\frac{\gamma}{\gamma \beta + 1} } \rvec{u} + \widetilde{\rvec{n}} }
\end{eqnarray}
where $\widetilde{\rvec{n}}$ is standard Gaussian noise. Since $\rvec{u}$ is a codewords from an optimal Gaussian codebook with power $1-\beta$, (\ref{appendix:eq:privateMessage}) was determined in \cite{EXIT}, and is,
\begin{align}
\lim_{\dim \to \infty} \InomGen{\rvec{u}}{\rvec{y}(\gamma)} & = \frac{1}{2} \log \left( 1 + \frac{\gamma ( 1 - \beta) }{\gamma \beta + 1} \right) \nonumber \\ &= \frac{1}{2} \log \left( \frac{1 + \gamma}{1 + \beta \gamma} \right)
\end{align}
for $\gamma \leq \snr_0$ (for $\gamma = \snr_0$ we have exactly the scalar Gaussian BC limit, thus we can see that without the assumption  on the private message acting as Gaussian i.i.d. noise, one could exceed this limit). For $\gamma > \snr_0$ the mutual information flattens and equals to the rate of the codebook.

Going on to the second term in (\ref{appendix:proofMIsuperposition}) we have:
\begin{eqnarray}
\Icond{\rvec{x}}{\rvec{y}(\gamma)}{\rvec{u}} = \Igen{\rvec{v}}{\sqrt{\gamma} \rvec{v} + \rvec{n} }
\end{eqnarray}
which is again the mutual information of an optimal Gaussian codebook, this time with power $\beta$,
\begin{eqnarray}
\lim_{\dim \to \infty} \frac{1}{\dim}\Icond{\rvec{x}}{\rvec{y}(\gamma)}{\rvec{u}} = \frac{1}{2} \log \left( {1 + \beta \gamma} \right).
\end{eqnarray}
This value remains valid for all $\gamma \leq \snr_1$. For $\gamma > \snr_1$ the above mutual information flattens and equals to the rate of this code. Adding the two terms together we obtain the desired expression (\ref{eq:thm_MIsuperposition}).

Now we turn to examine the derivative of $\I_{\dim}(\gamma)$ with respect to $\gamma$ (which is up to a factor of $\frac{1}{2}$ the $\MSEcoden(\gamma)$):
\begin{align}
\frac{\d}{\d \gamma} \I_{\dim}(\gamma) & =  \frac{\d}{\d \gamma} \frac{1}{\dim } \Igen{\rvec{u}}{\rvec{y}(\gamma)} + \frac{\d}{\d \gamma} \frac{1}{\dim }  \Icond{\rvec{x}}{\rvec{y}(\gamma)}{\rvec{u}} \nonumber \\
& = \frac{\d}{\d \gamma} \frac{1}{\dim } \Igen{\rvec{u}}{\sqrt{\tilde{\gamma} } \rvec{u} + \widetilde{\rvec{n}} } + \frac{\d}{\d \gamma} \frac{1}{\dim }  \Igen{\rvec{v}}{\sqrt{\gamma} \rvec{v} + \rvec{n} }
\end{align}
where $\tilde{\gamma} = \frac{\gamma}{\gamma \beta + 1}$. Examining the first expression on the right-hand-side we can use the chain rule. The derivative with respect to $\tilde{\gamma}$ is known \cite{EXIT} since we have an optimal Gaussian codebook of power $1-\beta$ transmitted over an additive Gaussian channel:
\begin{align}
\frac{\d}{\d \gamma} \frac{1}{\dim } \Igen{\rvec{u}}{\sqrt{\tilde{\gamma} } \rvec{u} + \widetilde{\rvec{n}} } & = \frac{\d}{\d \tilde{\gamma}} \frac{1}{\dim } \Igen{\rvec{u}}{\sqrt{\tilde{\gamma} } \rvec{u} + \widetilde{\rvec{n}} } \frac{\d}{\d \gamma} \frac{\gamma}{\gamma \beta + 1} \nonumber \\
& = \frac{1}{2} \frac{1 - \beta}{1 + \tilde{\gamma} (1 - \beta)}  \frac{1}{( 1 + \gamma \beta)^2} \nonumber \\
& = \frac{1}{2} \frac{1-\beta}{(1 + \gamma )(1 + \gamma \beta)}.
\end{align}
This is valid for $\gamma \leq  \snr_0$ after which the MMSE falls to zero.
The second expression on the right-hand-side is again an optimal Gaussian codebook of power $\beta$ transmitted over an additive Gaussian channel for which the derivative is the MMSE with known behavior \cite{EXIT}:
\begin{align}
\frac{\d}{\d \gamma} \frac{1}{\dim }  \Igen{\rvec{v}}{\sqrt{\gamma} \rvec{v} + \rvec{n} } & = \frac{1}{2} \frac{\beta}{1 + \gamma \beta}, \quad \gamma \leq \snr_1. \nonumber
\end{align}
At $\gamma = \snr_1$ the above expression falls to zero.
Putting the two together we obtain the desired result of equation (\ref{eq:thm:superpositionMSE}). This concludes the proof.
\end{IEEEproof}



\bibliographystyle{IEEEtran}

\bibliography{IEEEabrv,bib}

\end{document}